\documentclass[prl,aps,twocolumn,floatfix,showpacs]{revtex4}
\usepackage{graphicx,graphics,psfrag,amsmath,calc}
\usepackage{epsfig}
\usepackage{color}
\begin{document}

\title{Density fluctuations and compressibility matrix \\
for population or mass imbalanced Fermi-Fermi mixtures}

\author{Kangjun Seo and C. A. R. S{\'a} de Melo}
\affiliation{School of Physics, Georgia Institute of Technology, 
Atlanta, Georgia 30332, USA}
\date{\today}

\begin{abstract}
We describe the relation between the isothermal atomic compressibility and
density fluctuations in mixtures of two-component fermions with 
population or mass imbalance. We derive a generalized version of the
fluctuation-dissipation theorem which is valid for both balanced and 
imbalanced Fermi-Fermi mixtures. Furthermore, we show that
the compressibility, its critical exponents, and phase boundaries
can be extracted via an analysis of the density fluctuations 
as a function of population imbalance, interaction parameter or temperature.
Lastly, we demonstrate that in the presence of trapping potentials,
the local compressibility and local density-density correlations can 
be extracted via a generalized fluctuation-dissipation theorem
valid within the local density approximation.
\end{abstract}
\maketitle

%
%

Very recent experimental advances in Bose and Fermi systems 
have allowed for studies of density fluctuations and the use 
of the fluctuation-dissipation theorem
to obtain information about some thermodynamic properties 
of ultra-cold atoms. In the fermion case, the measurement 
of the density fluctuations and the atomic 
compressibility was extracted for non-interacting three-dimensional systems 
in harmonic traps~\cite{ketterle-10, esslinger-10}, 
while in the boson case, the connection between density fluctuations 
and compressibility was used to study superfluidity
in a two-dimensional system, and extract critical exponents associated with
the transition from a normal Bose gas to a Berezinskii-Kosterlitz-Thouless 
superfluid~\cite{chin-10}. The experimental extraction of the isothermal 
compressibility from density-fluctuation measurements 
were suggested several years ago both in harmonically 
confined systems~\cite{iskin-05} and optical lattices~\cite{iskin-06a}, 
but only recently improvements in the detection schemes of 
density fluctuations became sufficiently sensitive to extract 
this information from experimental 
data~\cite{ketterle-10, esslinger-10, chin-10}.

In principle, there are no 
major technical impediments to use the same technique
for the study of density fluctuations in population
imbalanced Fermi-Fermi mixtures 
of equal masses~\cite{ketterle-06, hulet-06}
or unequal masses~\cite{grimm-10}, where
the compressibility and spin susceptibility 
matrix elements could be directly extracted from the density
and density fluctuation profiles as discussed below.

In this manuscript, we show that 
some local and global thermodynamic properties can be extracted
through local measurements of densities and density fluctuations. 
We derive a generalized version of the fluctuation-dissipation theorem valid 
for any mixtures of atoms, and use it to analyze density fluctuations and
the compressibility of mixtures of two-component fermions with and without 
population imbalance at low temperatures.  For spatially uniform systems, 
we show that the global compressibility, phase boundaries
and critical exponents can be extracted from measurements 
of the density and density 
fluctuations for each component as a function of population imbalance,
interaction parameter or temperature.
While for spatially non-uniform systems, we show that the local 
and global compressibility 
can be extracted from measurements of the local density and local density 
fluctuations for each component.

%
%
{\it Hamiltonian:} To investigate the physics described above, we 
start with the real space Hamiltonian ($\hbar = 1$) density for three 
dimensional s-wave superfluids
\begin{equation}
\label{eqn:hamiltonian-real-space}
{\cal H} ({\bf r}) 
= 
\sum_{\alpha}
\psi^{\dagger}_{\alpha} ({\bf r}) 
\left( -\frac{\nabla^2}{2m_{\alpha}} - \mu_\alpha  
+ V_{\alpha} ({\bf r})\right) \psi_\alpha ({\bf r})
+ {\hat U} ({\bf r}),
\end{equation}
where 
$
{\hat U} ({\bf r}) =  
\int d{\bf r}^{\prime}  V_{\rm int} ( {\bf r}, {\bf r}^{\prime} ) 
\psi^{\dagger}_\uparrow ({\bf r}^{\prime})
\psi^{\dagger}_\downarrow ({\bf r}^{\prime})
\psi_{\downarrow}({\bf r})
\psi_{\uparrow} ({\bf r})
$
contains the interaction potential 
$
V_{\rm int} ({\bf r}, {\bf r}^{\prime}) 
= 
-g \delta ( {\bf r} - {\bf r}^{\prime} ),
$ 
and $\psi^{\dagger}_{\alpha} ({\bf r})$ creates  
fermions of mass $m_{\alpha}$ labeled by index $\alpha$. 
In addition, $V_{\alpha} ({\bf r})$  and $\mu_\alpha$ represent
the trapping potential and 
the chemical potential for each fermion type, respectively.
With the Hamiltonian
$ H = \int d {\bf r} {\cal H} ({\bf r})$,
we can study Fermi systems of equal masses 
$m_{\uparrow} = m_{\downarrow} = m$ with population imbalance,
or more generally we can study mixtures of fermions of 
unequal masses $m_{\uparrow} \ne m_{\downarrow}$.

From the grand partition function $Z = {\rm Tr} e^{-H/T}$, 
we can write the thermodynamic potential $\Omega = - T \ln Z$. 
First, we will ignore the trapping potential $V_{\alpha} ({\bf r})$ 
and discuss the spatially homogeneous case to simplify
the discussion, but will return to the spatially inhomogeneous
situation later. We set $V_{\alpha} ({\bf r})  = 0$ 
in Eq.~(\ref{eqn:hamiltonian-real-space}) 
and rewrite the Hamiltonian $H$ as
$H_1 - \sum_{\alpha} \mu_{\alpha} {\hat N}_{\alpha}$, where
$
H_1 
= 
\int 
d{\bf r}
\left[
\sum_{\alpha}
\psi^{\dagger}_{\alpha} ({\bf r}) 
\left( -\frac{\nabla^2}{ 2m_{\alpha} } 
\right) \psi_\alpha ({\bf r})
+ {\hat U} ({\bf r})
\right],
$
and 
$
{\hat N}_{\alpha}
=
\int 
d{\bf r}
\psi^{\dagger}_{\alpha} ({\bf r}) 
\psi_\alpha ({\bf r})
$
is the number operator for hyperfine state $\alpha$.
The average number of particles 
$N_{\alpha} = \langle {\hat N_\alpha} \rangle$ 
in hyperfine state $\alpha$, defined by the thermodynamic
average
$
\langle {\hat N}_{\alpha} \rangle 
= Z^{-1} 
{\rm Tr} 
\left[ {\hat N}_{\alpha} e^{-H/T}
\right],
$
can be rewritten in terms of the thermodynamic potential as
$ 
N_{\alpha} = 
\langle \hat N_{\alpha} \rangle 
= 
- \partial \Omega /\partial \mu_{\alpha} \vert_T.
$

{\it Pseudo-compressibility matrix:}
Next, we define the pseudo-compressibility matrix as
\begin{equation}
\label{eqn:pseudo-compressibility-matrix}
{\tilde \kappa}_{\alpha \beta} 
=
T 
\frac{\partial N_{\alpha} } {\partial \mu_{\beta} } \Big \vert_T
=
- 
T 
\frac{
\partial^2 \Omega }
{\partial \mu_{\alpha} \partial \mu_{\beta}} \Big \vert_T, 
\end{equation}
which through derivatives of $Z$ 
can be expressed as the thermodynamic average
$
{\tilde \kappa}_{\alpha \beta} 
=
\langle 
{\hat N}_{\alpha} 
{\hat N}_{\beta} 
\rangle
-
\langle 
{\hat N}_{\alpha} 
\rangle
\langle
{\hat N}_{\beta} 
\rangle.
$
The mechanical stability of the system is guaranteed when
both eigenvalues of $\tilde \kappa_{\alpha \beta}$ are positive definite.
Furthermore, ${\tilde \kappa}_{\alpha \beta}$ is a measure 
of density-density fluctuations:
\begin{equation}
\label{eqn:fluctuation-dissipation}
{\tilde \kappa}_{\alpha\beta} 
= 
\langle
(
\hat N_{\alpha} - N_{\alpha}
)
(
\hat N_{\beta} - N_{\beta} 
)
\rangle .
\end{equation}
The corresponding generalized compressibility matrix 
$\kappa_{\alpha \beta}$ can be obtained from 
$\tilde \kappa_{\alpha \beta}$ through the relation
$
\kappa_{\alpha \beta} 
= 
{\tilde \kappa}_{\alpha \beta}  
/
\left[
\langle {\hat N}_{\alpha} \rangle
\langle {\hat N}_{\beta} \rangle 
\right],
$
describing a generalized fluctuation-dissipation theorem for 
multicomponent fermions. 

Using similar experimental techniques 
to those described in Refs.~\cite{ketterle-10, esslinger-10},
it may be possible to measure the matrix 
elements of $\tilde \kappa_{\alpha \beta}$ directly.
Thus, it is important to identify the relation between the
isothermal compressibility 
$\kappa_T^{-1} 
= - V (\partial P / \partial V) \vert_T$ and the elements
of the compressibility matrix $\kappa_{\alpha \beta}$.

{\it Isothermal compressibility:}
The relation between $\kappa_T$ and $\kappa_{\alpha \beta}$ can be
established by recalling that the thermodynamic potential 
$\Omega = -PV$, where $P$ is the pressure and $V$ is the volume of the 
system. Defining $G = \Omega + PV = 0$, and recalling that
$\Omega$ is a function of 
temperature $T$, volume $V$ and chemical potentials 
$\mu_{\alpha}$ results in
$
dG 
= 
-SdT 
-
N_\uparrow d \mu_\uparrow
- 
N_\downarrow d \mu_\downarrow 
+ 
V dP 
= 
0.
$ 
At constant temperature
$dT = 0$, we can establish the relation 
$
- V dP\vert_T 
= 
N_\uparrow d \mu_\uparrow \vert_T
+ 
N_\downarrow d \mu_\downarrow \vert_T.
$

This means that the inverse isothermal compressibility 
$
\kappa_T^{-1} 
= 
- V (\partial P / \partial V)_T
$ 
can be writen in terms of isothermal partial derivatives
of $\mu_{\alpha}$ with respect to volume 
$
\kappa_T^{-1} 
= 
N_\uparrow (\partial \mu_\uparrow / \partial V)\vert_T
+
N_\downarrow (\partial \mu_\downarrow / \partial V)\vert_T.
$
But in turn the partial derivatives 
$
\partial \mu_{\alpha} 
/ 
\partial V \vert_T
$ 
can be expressed in terms
$\partial \mu_{\alpha}/\partial N_{\beta}\vert_T$ 
and $N_\beta$, leading to 
\begin{equation}
\label{eqn:isothermal-compressibility}
\frac{1}{\kappa_T}
=
\frac{T}{V}
\left[
\frac{N_\uparrow^2}{\tilde \kappa_{\uparrow \uparrow}}
+
\frac{N_\uparrow N_\downarrow}{\tilde \kappa_{\uparrow \downarrow}}
+
\frac{N_\downarrow N_\uparrow}{\tilde \kappa_{\downarrow \uparrow}}
+
\frac{N_\downarrow^2}{\tilde \kappa_{\downarrow \downarrow}}
\right].
\end{equation}
This expression can be written in the compact form $V T^{-1} \kappa_T^{-1}
= \sum_{\alpha \beta} \left[ \kappa_{\alpha \beta} \right]^{-1}$, 
by using the definition 
$\kappa_{\alpha \beta} = \tilde \kappa_{\alpha \beta}/(N_\alpha N_\beta)$.
Therefore direct measurements of $\kappa_{\alpha \beta}$ lead
to the isothermal compressibility $\kappa_T$ of the system.

{\it Connection to pseudo-spin susceptibility:}
We can also work with the total number of
particles $N_+ =  N_\uparrow + N_\downarrow$, the
particle number difference $N_- = N_\uparrow - N_\downarrow$,
and their corresponding chemical potentials
$\mu_\pm = ( \mu_\uparrow \pm \mu_\downarrow )/2 $, respectively.
In this case, we can define a similar 
pseudo-compressibility tensor 
$
{\tilde \kappa}_{ij} 
=
T \partial \langle {\hat N}_i \rangle / \partial \mu_j \vert_T
=
- 
T \partial^2 \Omega 
/ 
\partial \mu_{i} \partial \mu_{j} \vert_T, 
$
leading to 
$
{\tilde \kappa}_{ij} 
=
\langle 
{\hat N}_i 
{\hat N}_j
\rangle
-
\langle 
{\hat N}_i 
\rangle
\langle
{\hat N}_j 
\rangle,
$
where the indices $i,j$ can each take $\pm$ values.
The corresponding expression for $\kappa_T$
has exactly the same form as before:
\begin{equation}
\label{eqn:isothermal-compressibility-pm}
\frac{1}{\kappa_T}
=
\frac{T}{V}
\left[
\frac{N_+^2}{\tilde \kappa_{+ +}}
+
\frac{N_+ N_-}{\tilde \kappa_{+ -}}
+
\frac{N_-  N_+}{\tilde \kappa_{- +}}
+
\frac{N_-^2}{\tilde \kappa_{- -}}
\right].
\end{equation}
It is clear that $\kappa_T$ reduces to the standard result 
for balanced populations where $N_- = 0$ and $N_+ = N$,
$\kappa_T^{-1} = T V^{-1} N_+^2/\tilde \kappa_{++}$, 
leading to the standard form of the fluctuation-dissipation
theorem:
$
\kappa_T 
= 
V T^{-1} 
\left[ 
\langle {\hat N}^2 \rangle
-\langle {\hat N} \rangle^2
\right]/ \langle {\hat N} \rangle^2.
$
The connection between the two representations is simple.
The first diagonal term is 
$
{\tilde \kappa}_{++} = 
{\tilde \kappa}_{\uparrow \uparrow}
+
{\tilde \kappa}_{\downarrow \downarrow}
+
2{\tilde \kappa}_{\uparrow \downarrow},
$
the second diagonal term is
$
{\tilde \kappa}_{--} = 
{\tilde \kappa}_{\uparrow \uparrow}
+
{\tilde \kappa}_{\downarrow \downarrow}
-
2{\tilde \kappa}_{\uparrow \downarrow},
$
while the off-diagonal terms 
$
{\tilde \kappa}_{+-} 
=
{\tilde \kappa}_{-+}
=
{\tilde \kappa}_{\uparrow \uparrow}
-
{\tilde \kappa}_{\downarrow \downarrow}
$  
are identical by symmetry.

The pseudo-spin susceptibility $\chi_{zz}$ of the system can 
also be extracted from the pseudo-compressibility matrix 
$\tilde \kappa_{ij}$, since $\mu_{-}$ plays
the role of an effective magnetic field $h_z$ along the quantization axis
$z$, and $N_{-}$ plays the role of the magnetization $m_z$. For imbalanced
Fermi systems this implies that 
\begin{equation}
\label{eqn:pseudo-spin-susceptibility}
\tilde \kappa_{--} 
=  
T 
\frac{ \partial  N_{-} } {\partial \mu_{-} } \Big\vert_T
=
T
\frac{ \partial m_z } { \partial h_z } \Big\vert_T
=
T\chi_{zz},
\end{equation}
and generalizes the results obtained for ultra-cold fermions interacting via 
$p$-wave interactions~\cite{botelho-05}.

{\it Compressibility matrix in a trap:}
In the presence of a trapping potential $V_{\alpha} ({\bf r})$,
we define the local chemical potential as
$
\mu_\alpha ({\bf r}) 
= 
\mu_{\alpha} 
- 
V_{\alpha} ({\bf r}),
$ 
and rewrite the Hamiltonian explicitly as
$
H 
=
H_1 
-
\int d{\bf r} 
\mu_{\alpha} ({\bf r}) 
{\hat n}_{\alpha} ({\bf r}),
$
where 
$
{\hat n}_{\alpha} ({\bf r}) 
= 
\psi^{\dagger} ({\bf r}) \psi ({\bf r})
$
corresponds to the particle density operator.
In this case, the grand partition function is  
$
Z 
\left[
T, V, \mu_{\alpha} ({\bf r})
\right]
=
{\rm Tr}
e^{
- 
\left[
H_1 
- 
\int
d{\bf r}
\mu_{\alpha} 
{\hat n}_{\alpha} ( {\bf r} )
\right]
/
T
}.
$
The thermodynamic potential 
$
\Omega 
\left[
T, V, \mu_{\alpha} ({\bf r}) 
\right]
$
is also a functional of $\mu_{\alpha} ({\bf r})$, 
such that local and
even non-local quantities can be 
extracted. For instance the local density 
$
n_{\alpha} ({\bf r}) 
= 
\langle 
{\hat n}_{\alpha} ({\bf r}) 
\rangle
$ 
is simply written as the functional derivative
$
V n_{\alpha} ({\bf r}) 
=
N_{\alpha} ({\bf r})
= 
-\delta \Omega / \delta \mu_{\alpha} ({\bf r}),
$
where $N_{\alpha} ({\bf r})$ is the local
number of particles.
Correspondingly the non-local
pseudo-compressibility matrix is 
$
\tilde \kappa_{\alpha \beta} ( {\bf r}, {\bf r}^\prime )
=
-T \delta^2 \Omega 
/ 
\delta \mu_{\alpha} ({\bf r}) \delta \mu_{\beta} ({\bf r}^\prime ),
$
which in terms of the local particle number operators 
${\hat N}_{\alpha} ({\bf r})$ and
${\hat N}_{\beta} ({\bf r}^\prime)$  
becomes
\begin{equation}
\tilde \kappa_{\alpha \beta} ( {\bf r}, {\bf r}^\prime )
=
\langle
{\hat N}_{\alpha} ({\bf r}) {\hat N}_{\beta} ({\bf r}^\prime)
\rangle
- 
\langle
{\hat N}_{\alpha} ({\bf r})
\rangle
\langle
{\hat N}_{\beta} ({\bf r}^\prime)
\rangle.
\end{equation}
Since 
$
N_{\alpha} ({\bf r}) 
= 
\langle {\hat N}_{\alpha} ({\bf r})
\rangle
$
is a local thermodynamic average, we rewrite 
$
\tilde \kappa_{\alpha \beta} ({\bf r}, {\bf r}^\prime)
= 
\langle \delta {\hat N}_{\alpha} ({\bf r}) 
\delta {\hat N}_{\beta} ({\bf r}^\prime) \rangle,
$
where 
$
\delta {\hat N}_{\alpha} ({\bf r}) 
= 
{\hat N}_{\alpha} ({\bf r})
-
N_{\alpha} ({\bf r})
$
is the local fluctuation in the number of particles 
of type $\alpha$.
The local pseudo-compressibility matrix is simply 
$
\tilde \kappa_{\alpha \beta} ({\bf r}) 
=
{\tilde \kappa}_{\alpha \beta} ( {\bf r}, {\bf r} ).
$

Within the local density approximation (LDA) the
local isothermal compressibility $\kappa_T ({\bf r})$
can be derived from the local pressure
$ P({\bf r}) $ as 
$
\kappa_T ({\bf r}) 
= 
- V \partial P ({\bf r})/ \partial V \vert_T.
$ 
Following the steps leading to 
Eq.~(\ref{eqn:isothermal-compressibility}), we obtain 
\begin{equation}
\frac{1}{ \kappa_T ({\bf r}) }
= 
\frac{T}{V}
\sum_{\alpha \beta} 
\frac{1}{\kappa_{\alpha \beta} ({\bf r})},
\end{equation}
which is the local generalization of the fluctuation-dissipation
theorem within LDA. Here, 
$
\kappa_{\alpha \beta} ({\bf r}) 
=
{\tilde \kappa}_{\alpha \beta} ({\bf r})
/
\left[
N_{\alpha} ({\bf r}) N_{\beta} ({\bf r})  
\right].
$ 

%
%
{\it Imbalanced Fermi-Fermi mixtures:}
As an example of the general relations just derived 
we discuss the case of imbalanced Fermi-Fermi mixtures with 
equal masses, which has attracted a lot 
of interest~\cite{yip-06, mueller-06, iskin-06b, stoof-06,
pieri-06, tempere-07}.

The resulting action corresponding to the 
Hamiltonian given in Eq.~(\ref{eqn:hamiltonian-real-space})
has been succesfully calculated for the case of a uniform 
superfluid~\cite{iskin-07} in the Gaussian approximation
as
$
S_{G} 
= 
S_0 
+ 
\frac{1}{2T}
\sum_{q} 
\Lambda^\dagger (q)
{\bf F}^{-1}
\Lambda (q),
$
where $q = ({\bf q}, \nu_{\ell} )$, with 
$\nu_{\ell} = 2\pi \ell T$ being the 
bosonic Matsubara frequeny at temperature $T$.
Here, $\Lambda (q)$ is the order parameter fluctuation
field and the matrix ${\bf F}^{-1} (q)$ is the inverse
fluctuation propagator. Furthermore,
$$
S_0 
=
\frac{\vert \Delta_0 \vert^2}{g T}
+
\frac{1}{T}
\sum_{\bf k} 
\left(
\xi_{ {\bf k}, + } - E_{ {\bf k}, + }
\right)
+ 
\sum_{\alpha}
\ln 
\left[
n_F 
(
- E_{ {\bf k}, \alpha }
)
\right],
$$
is the saddle point action, 
where 
$
E_{ {\bf k}, \alpha } 
=
\sqrt{
\xi_{{\bf k}, +}^2 + \vert \Delta_{\bf k} \vert^2
}
+ s_{\alpha} \xi_{ {\bf k}, -} 
$
is the quasiparticle energy when 
$s_{\uparrow} = 1$ and is the negative of the
quasihole energy when $s_{\downarrow} = -1$.
We also use the notation 
$
E_{{\bf k}, \pm} 
=
\left(
E_{{\bf k}, \uparrow} \pm E_{{\bf k}, \downarrow}
\right)/2.
$
In addition, $\Delta_{\bf k} = \Delta_0 \Gamma_{\bf k}$
is the order parameter for superfluidity for pairing 
with zero center of mass momentum, $\Gamma_{\bf k} = 1$
for s-wave pairing,  
$n_F ( E_{ {\bf k}, \alpha } )$ is the Fermi distribution,
and 
$
\xi_{{\bf k}, \pm}
= 
\left(
\xi_{{\bf k}, \uparrow} \pm 
\xi_{{\bf k}, \downarrow} 
\right)/2
= 
k^2/(2m_{\pm})-\mu_\pm,
$
where  
$
m_{\pm} 
= 
2 m_{\uparrow} m_{\downarrow}
/
\left(
m_{\downarrow} \pm m_{\uparrow}
\right)
$.
Notice that $m_{+}$ is twice the reduced mass
of the $\uparrow$ and $\downarrow$ fermions, and that
the equal mass case $( m_{\uparrow} = m_{\downarrow} )$ 
corresponds to $\vert m_{-} \vert \to \infty$.

The fluctuation term in the action leads to 
a correction to the thermodynamic potential, 
which can be written as 
$
\Omega_{G} 
= 
\Omega_0 
+ 
\Omega_{\rm fluct},
$
where $\Omega_0 = T S_0$
and 
$
\Omega_{\rm fluct} 
= 
T
\sum_{q}
\ln det 
\left[
T {\bf F}^{-1} (q)
\right].
$
The saddle point condition 
$
\delta\Omega_0/\delta \Delta_0^*
=
0
$
leads to the order parameter equation
\begin{equation}
\label{eqn:order-parameter}
\frac{1}{g}
= 
\sum_{\bf k}
\frac{ \vert \Gamma_{\bf k} \vert^2 } { 2 E_{ {\bf k}, + } }
X_{ {\bf k}, + },
\end{equation}
where 
$
X_{ {\bf k}, \pm }
=
\left(
X_{ {\bf k}, \uparrow }
\pm
X_{ {\bf k}, \downarrow }
\right)/2,
$
with 
$
X_{ {\bf k}, \alpha}
=
\tanh 
\left(
E_{ {\bf k}, \alpha } / (2T)
\right).
$
As usual, we eliminate $g$ in favor
of the scattering length $a_s$ via the
relation 
$
1/g 
= 
- m_{+} V/(4\pi a_s) 
+ 
\sum_{\bf k}
\vert \Gamma_{\bf k} \vert^2
/
(2 \epsilon_{ {\bf k}, +} ),
$
where 
$
\epsilon_{ {\bf k}, \pm }
=
\left(
\epsilon_{ {\bf k}, \uparrow }
\pm 
\epsilon_{ {\bf k}, \downarrow }
\right)/2 
=
k^2/(2m_\pm)
$.
The order parameter equation needs to be solved 
self-consistently with the number equations
$
N_{\alpha} 
= 
- \frac{ \partial \Omega } { \partial \mu_{\alpha} }
\Big \vert_T,
$
which has two contributions
\begin{equation}
\label{eqn:number}
N_{\alpha} 
=
N_{0, \alpha}
+
N_{{\rm fluct}, \alpha},
\end{equation}
where 
$
N_{0, \alpha}=
-\frac{\partial \Omega_0}{ \partial \mu_{\alpha} } \Big \vert_T
$ is the saddle point
number equation given by
\begin{equation}
N_{0, \alpha} 
=
\sum_{\bf k}
\left(
\frac{1 - s_{\alpha} X_{{\bf k}, - }}{2}
-
\frac{\xi_{ {\bf k}, + } } {2 E_{{\bf k}, +} }  X_{{\bf k}, +} 
\right)
\end{equation}
and 
$
N_{{\rm fluct}, \alpha}
=
- 
\partial \Omega_{\rm fluct} / \partial \mu_{\alpha}\vert_T
$
is the fluctuation contribution to $N_{\alpha}$
given by 
$
N_{{\rm fluct}, \alpha }
=
-
T
\sum_{q}
\{
\partial
\left[
det {\bf F}^{-1} (q) 
\right]
/
\partial \mu_{\alpha}
\}
/
det {\bf F}^{-1} (q).
$

To calculate ${\tilde \kappa}_{\alpha \beta}$ and $\kappa_T$ from the
thermodynamic potential $\Omega$, we note that 
$
\Omega 
\left[
\Delta_0 (\mu_{\uparrow}, \mu_{\downarrow}),
\mu_{\uparrow}, \mu_{\downarrow}, T
\right]
\to
\Omega 
\left[
\mu_{\uparrow}, \mu_{\downarrow}, T
\right].
$
Because of the implicit dependence of $\Delta_0$ on $\mu_{\uparrow}$
and $\mu_{\downarrow}$, the calculation of ${\tilde \kappa}_{\alpha \beta}$
requires
\begin{equation}
{\tilde \kappa}_{\alpha \beta}
=
T 
\frac{\partial N_{\alpha}} {\partial \mu_{\beta} } \Big \vert_{T, e}
+
T \frac{\partial N_{\alpha}} {\partial \vert \Delta_0 \vert^2} 
\Big \vert_{T, e}
\cdot
\frac{\partial \vert \Delta_0 \vert^2} {\partial \mu_\beta} 
\Big \vert_{T,i},
\end{equation}
where the label ``$e$'' (``$i$'') means explicit (implicit) derivative.
The mechanical stability of the uniform superfluid and normal
phases is guaranteed if all eigenvalues of $\tilde \kappa_{\alpha \beta}$
(or $\tilde \kappa_{ij}$) are positive. 
This is achieved when ${\rm Tr} \tilde \kappa > 0$ and
${\rm det} \tilde \kappa > 0$. When the lowest eigenvalue of 
${\tilde \kappa}$ reaches zero then the system becomes 
mechanically unstable. 

\begin{figure} [htb]
\centering
\begin{tabular}{cc}
\epsfig{file = 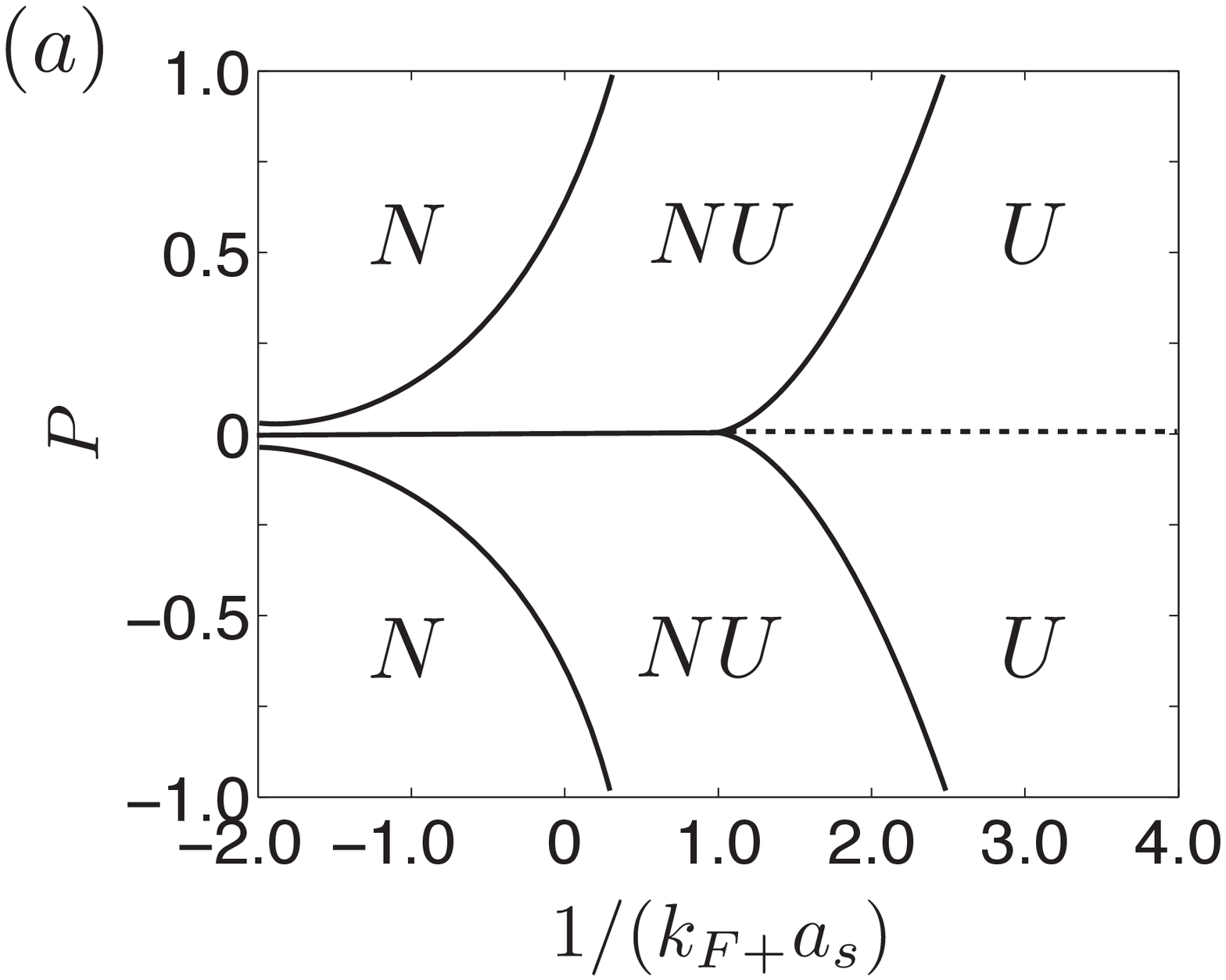, width=0.5\linewidth,clip=} &
\epsfig{file = 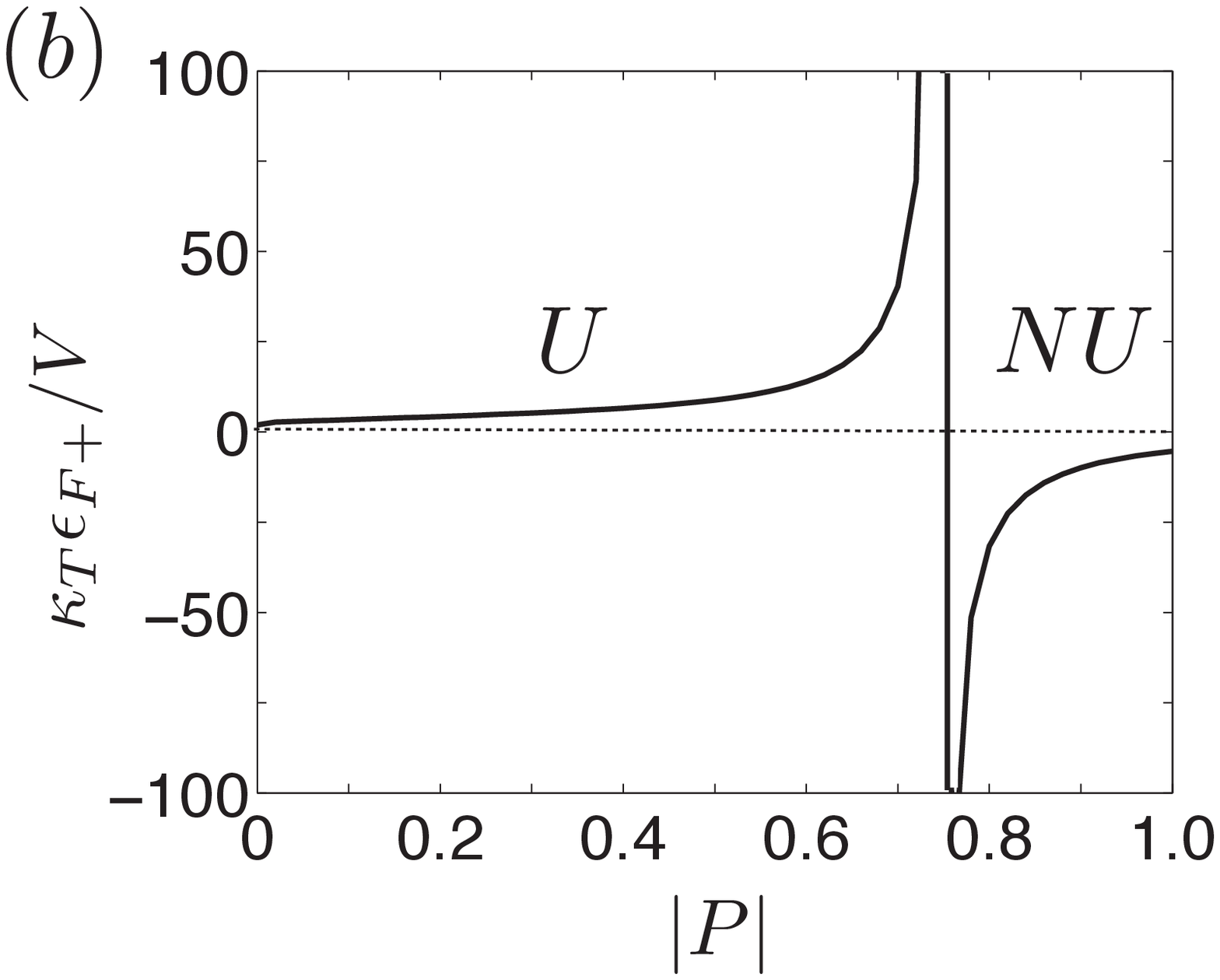, width=0.5\linewidth,clip=} 
\end{tabular}
\caption{ 
\label{fig:one}
a) The zero temperature phase diagram of population imbalance 
$P = N_{-}/N_{+}$ versus scattering parameter $1/( k_{F +} a_s )$
for equal mass fermions is shown.
b) The dimensionless isothermal compressibility $\kappa_T \epsilon_{F_+}/V$ is
shown for fixed $1/(k_{F +} a_s) = 2.16$.
}
\end{figure}

We define the Fermi momentum 
$k_{F +}^3 = k_{F \uparrow}^3 + k_{F \downarrow}^3$,
where $k_{F \alpha}$ is the Fermi momentum of
each species, and the Fermi energy 
$\epsilon_{F+} = k_{F +}^2/(2m_{+})$.
For a mixture of fermions of equal masses, different hyperfine states
and no trapping potential, the zero temperature phase diagram of 
population imbalance 
$
P = ( N_{\uparrow} - N_{\downarrow} ) /  ( N_{\uparrow} +  N_{\downarrow} )
= 
N_{-} / N_{+}
$ 
versus scattering parameter $ 1/( k_{F +} a_s ) $
is shown in Fig.~\ref{fig:one}a, 
where the normal (N), non-uniform (NU) and uniform (U)
superfluid regions are indicated.
The isothermal compressibility $\kappa_T$ is shown in
Fig.~\ref{fig:one}b for $1/(k_{F_+} a_s) = 2.16$ and
changing $P$. Notice that as $P$ increases,
the dimensionless compressibility $\kappa_T \epsilon_{F_+}/ V$ diverges at
a critical population imbalance $P_c = 0.77$, and becomes negative
for $P > P_c$ signaling a quantum phase transition from uniform
superfluidity with coexistence of excess unbound fermions and 
paired fermions in the same spatial region to a phase separated
regime where excess unbound fermions and paired fermions tend
to avoid being in the same region of space.

\begin{figure} [htb]
\centering
\begin{tabular}{cc}
\epsfig{file = 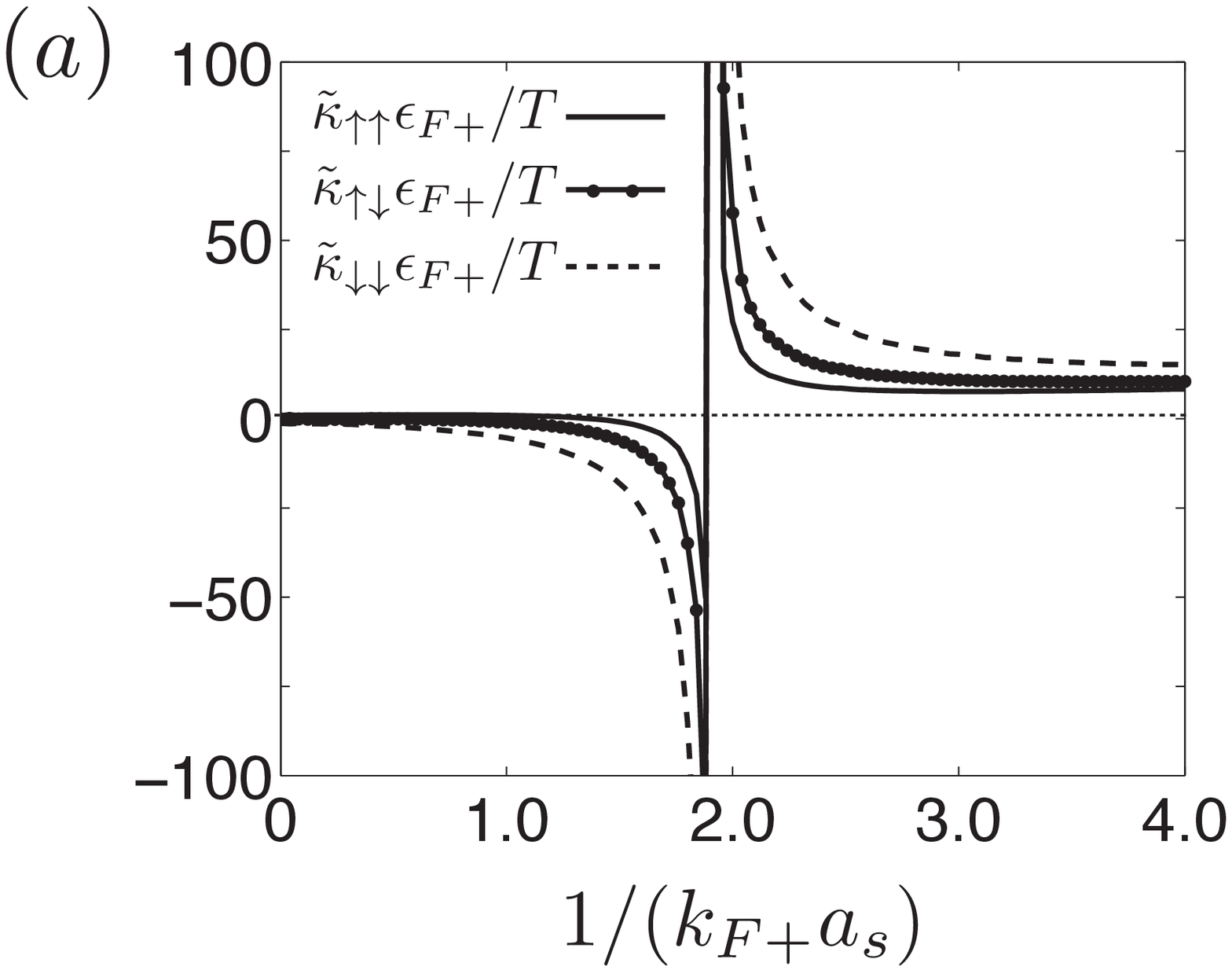, width=0.5\linewidth,clip=} &
\epsfig{file = 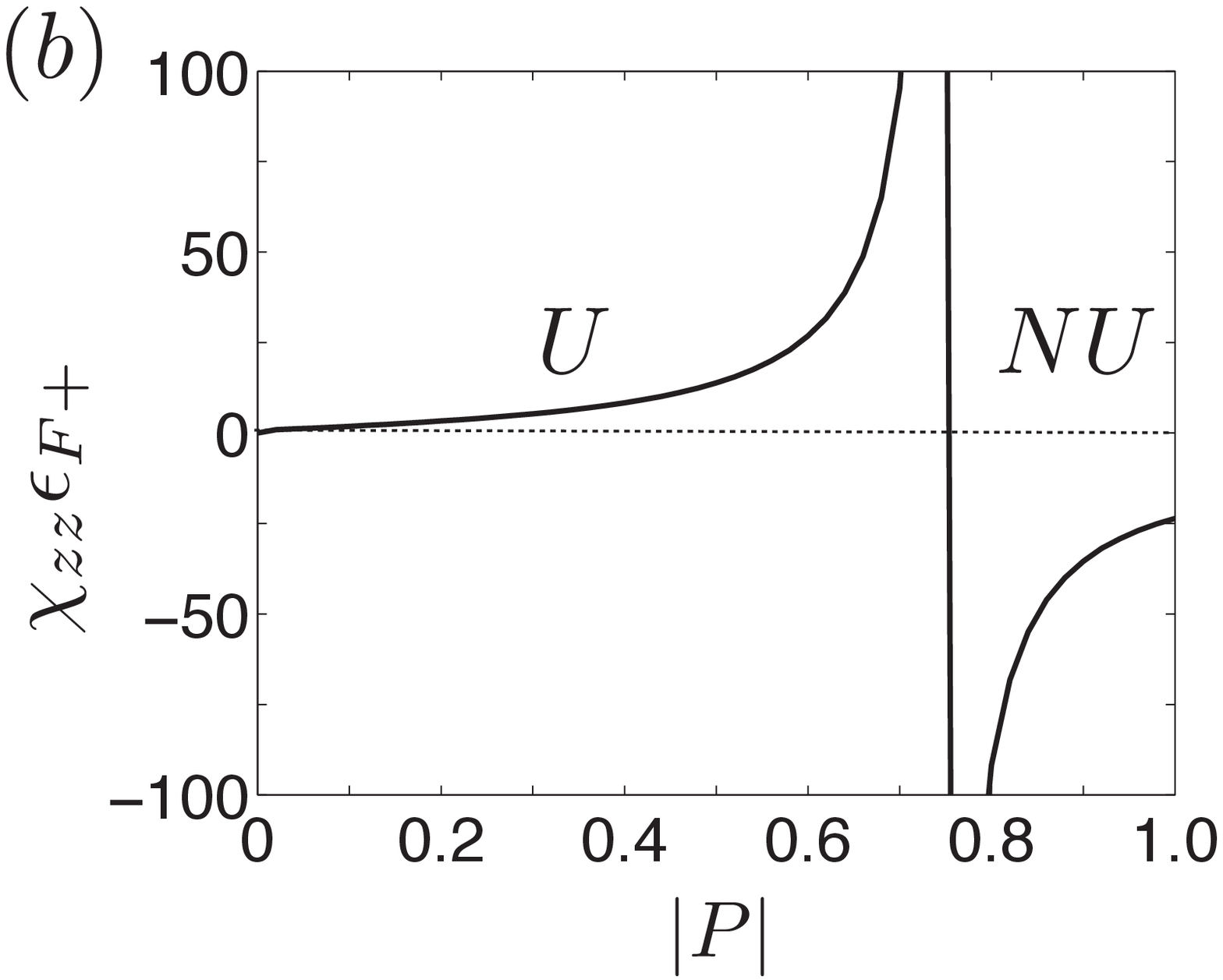, width=0.5\linewidth,clip=} 
\end{tabular}
\caption{ 
\label{fig:two}
a) Pseudo-compressibility matrix elements 
${\tilde \kappa}_{\alpha \beta}\epsilon_{F+}/T$
as a function of $1/(k_{F +} a_s) $ for population imbalance $P = 0.5$;
b) Pseudo-spin susceptibility 
$\chi_{zz}\epsilon_{F +} = \tilde \kappa_{--} \epsilon_{F +}/T$
as a function of $P$ for interaction
parameter $1/(k_{F +} a_s) = 2.16$.
}
\end{figure}

In Fig.~\ref{fig:two}a, we show the 
dimensionless matrix elements 
${\tilde \kappa}_{\alpha \beta}\epsilon_{F +}/T$
as a function of $1/(k_{F_+} a_s)$ on the BEC side 
$[ 1/(k_{F_+} a_s) > 0 ]$ for fixed population 
imbalance $P = 0.5$.
Notice that ${\tilde \kappa}_{\alpha \beta}$
diverges as 
$
( \lambda - \lambda_c )^{-1},
$
where 
$
\lambda 
= 
1/(k_{F_+} a_s),
$
and $\lambda_c = 1.9$ is the critical interaction
parameter. 
As seen in Fig.~\ref{fig:two}b, 
when population imbalance is changed in the BEC regime, 
e.g. $1/(k_{F_+} a_s ) = 2.16$, the
dimensionless pseudo-spin susceptibility 
$\chi_{zz} \epsilon_{F+} = {\tilde \kappa}_{--} \epsilon_{F+}/T$ 
diverges at the phase boundary between the uniform superfluid
and the non-uniform phases as 
$\chi_{zz} \epsilon_{F+} \sim (P - P_c)^{-1}$. 
The negative values of the 
matrix elements ${\tilde \kappa}_{\alpha \beta}\epsilon_{F +}/T$ 
and $\chi_{zz} \epsilon_{F+}$ just indicate the region of non-uniform superfluidity, 
i.e., the region where uniform superfluidity is not mechanically stable. 
Thus, the  generalized fluctuation-dissipation theorem
described in Eq.~(\ref{eqn:fluctuation-dissipation}) 
allows for the extraction of critical exponents of density-density
and pseudospin-pseudospin correlations accross phase boundaries.

In the case of a non-zero trapping potential 
$V_{\alpha} ({\bf r}),$ 
we make use of LDA
and obtain the local thermodynamic potential 
$
\Omega ({\bf r}) 
= 
\Omega 
\left[ 
\mu_{\uparrow} ({\bf r}), \mu_{\downarrow} ({\bf r})
\right]
$
from the thermodynamic potential in the absence of 
a trap 
$
\Omega 
\left[ 
\mu_{\uparrow}, \mu_{\downarrow}
\right],
$
via the substitution 
$
\mu_{\alpha} 
\to 
\mu_{\alpha} ({\bf r}) 
= 
\mu_{\alpha} - V_{\alpha} ({\bf r}).
$ 
This implies that in Eqs.~(\ref{eqn:order-parameter}) 
and~(\ref{eqn:number})
the order parameter $\Delta_0$ and
the number of particles $N_{\alpha}$ become functions
of position ${\bf r}$ via the position dependent chemical 
potentials $\mu_{\alpha} ({\bf r})$. 
As a result we have 
$
\Delta_0 ({\bf r}) 
= 
\Delta_0 
\left[
\mu_{\uparrow} ( {\bf r} ), 
\mu_{\downarrow} ( {\bf r} )
\right]
$
and
$
N_{\alpha} ({\bf r})
=
N_{\alpha}
\left[
\mu_{\uparrow} ( {\bf r} ), 
\mu_{\downarrow} ( {\bf r} )
\right].
$

\begin{figure} [htb]
\centering
\begin{tabular}{cc}
\epsfig{file = 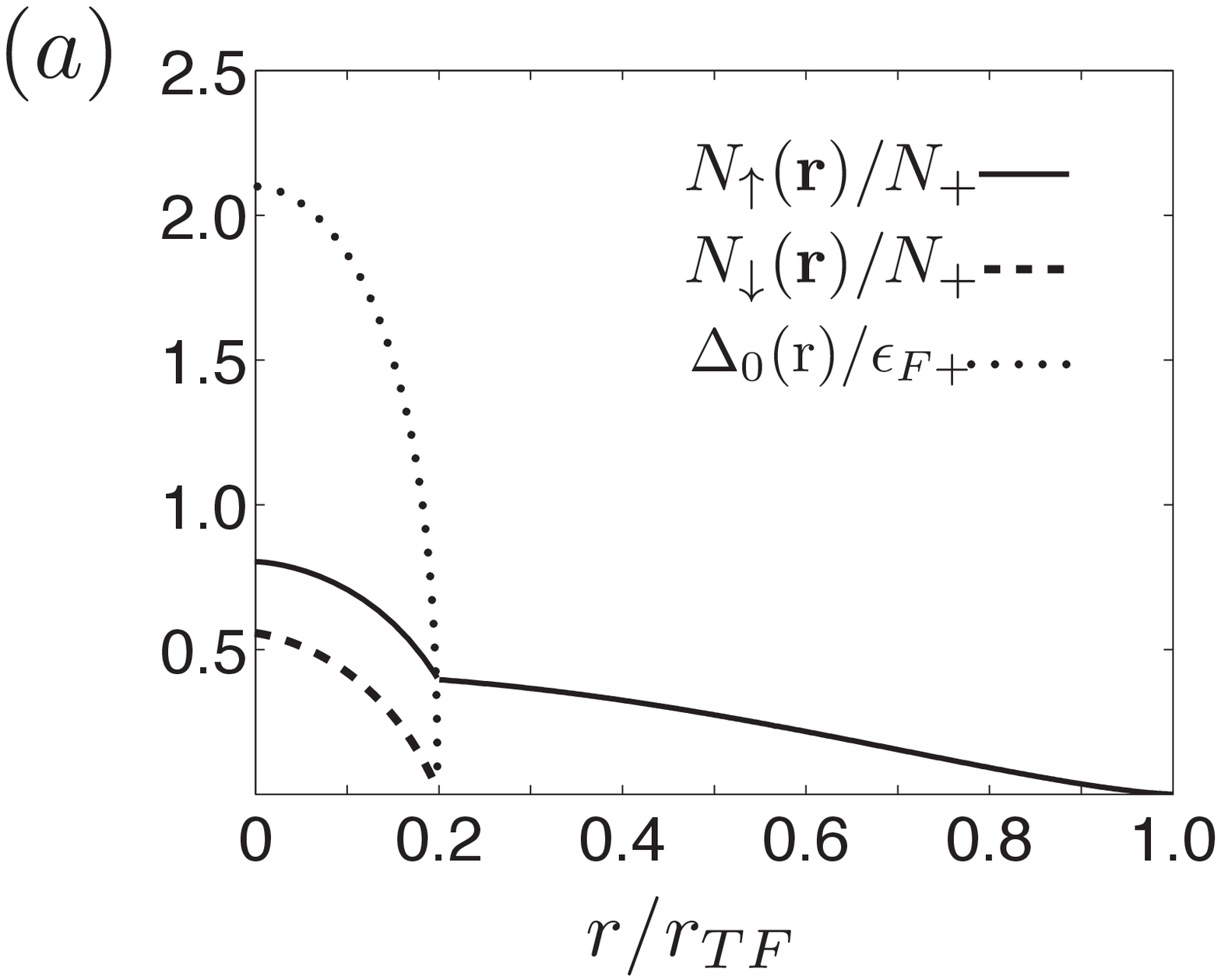, width=0.5\linewidth,clip=} &
\epsfig{file = 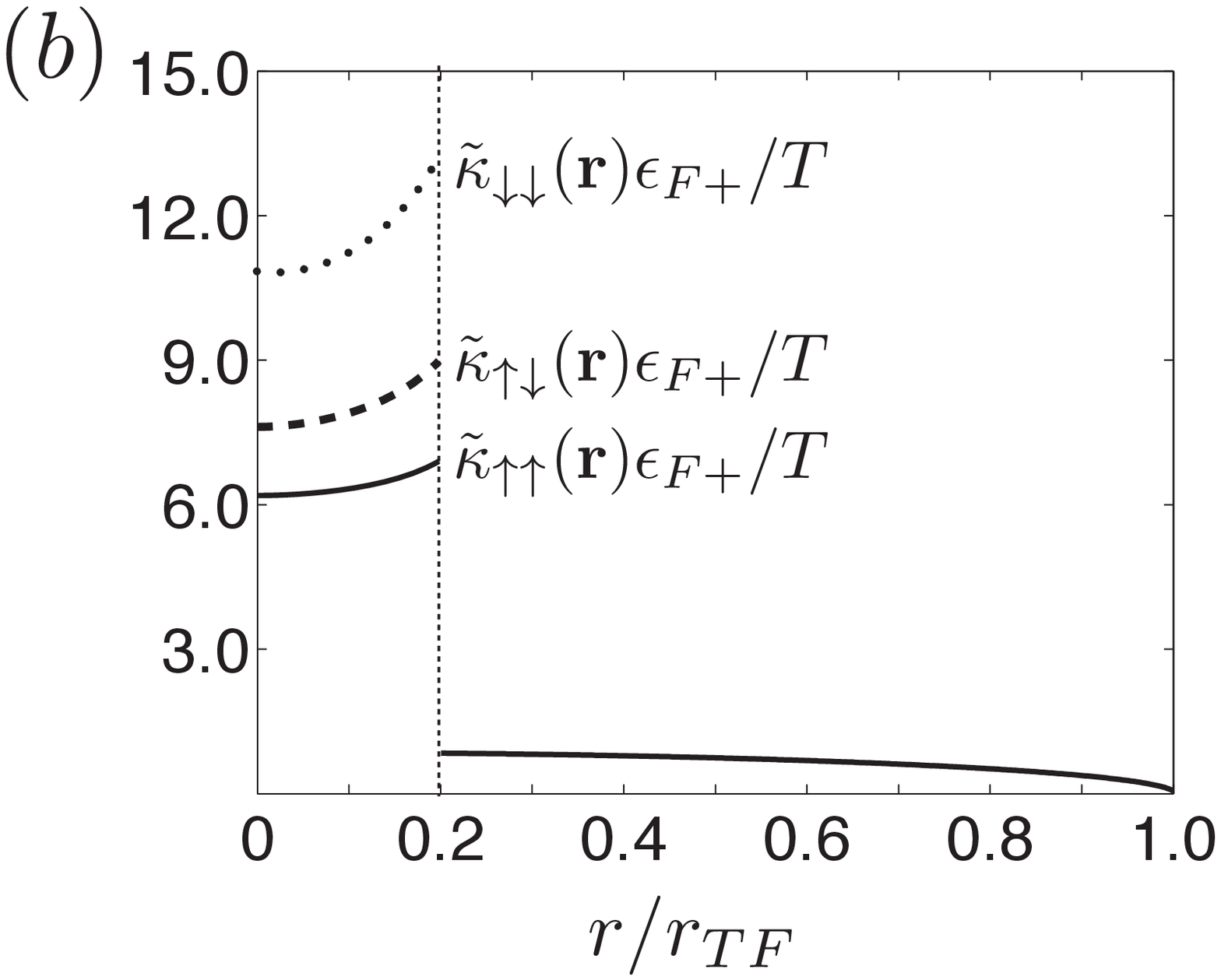, width=0.5\linewidth,clip=} 
\end{tabular}
\caption{ 
\label{fig:three}
a) Particle number profiles $N_{\alpha}({\bf r})/N_+$,
and order parameter $\Delta_0 ({\bf r})/\epsilon_{F +}$;
b) Matrix elements 
${\tilde \kappa}_{\alpha \beta} \epsilon_{F +}/T$ 
as a function of ${\bf r}/r_{TF}$,
for population imbalance $P = 0.6$ and interaction
parameter $1/(k_{F +} a_s) = 3.0$. 
}
\end{figure}

We consider harmonic trapping potentials
$V_{\alpha} ({\bf r}) = \gamma_\alpha r^2/2$,
where $\gamma_{\alpha} = m_{\alpha} \omega_{\alpha}$,
with $\omega_{\alpha}$ being the trapping frequencies 
of fermion of type $\alpha$.
For the equal mass case, we show in 
Fig.~\ref{fig:three}a,
the particle number
profiles $N_{\alpha} ({\bf r})$ and
the order parameter $\Delta_0 ({\bf r})$ 
as a function of dimensionless position
${\bf r}/r_{TF}$, where $r_{TF}$ is the Thomas-Fermi
radius defined through the condition
$
\epsilon_{F +} = 
\gamma_{+} r_{TF}^2/2,
$
where 
$
\gamma_{+} 
= 
\gamma_{\uparrow} + \gamma_{\downarrow}.
$
These spatial profiles show that superfluidity coexists
with excess unpaired fermions, but the majority of excess
unpaired fermions are pushed away from the center of the 
trap. In Fig.~\ref{fig:three}b, we show 
the spatial dependence of 
${\tilde \kappa}_{\alpha \beta} ({\bf r})$, 
from which the local correlation functions 
$\langle {\hat N}_{\alpha} ({\bf r}) {\hat N}_{\beta} ({\bf r}) \rangle$
and the pseudo-spin susceptibility $\chi_{zz} ({\bf r}) = {\tilde \kappa}_{--}/T$ 
can also be easily extracted.
Within LDA, ${\tilde \kappa}_{\alpha \beta} ({\bf r})$ exhibit a discontinuous 
jump at the position ${\bf r}_c$ where $\Delta_0 ({\bf r}_c) = 0$.
In the superfluid region, local particle fluctuations 
reveal the extra correlations brought in by full pairing such that
$\chi_{zz} ({\bf r}) = 0$. 
Outside the superfluid region local particle fluctuations show a
decrease in particle-particle correlations, as the excess
unpaired fermions are pushed away from the center of the trap,
leading to $\chi_{zz} ({\bf r}) \ne 0$.

{\it Summary:}
We derived a generalized fluctuation-dissipation theorem 
for Fermi-Fermi mixtures, which was used to extract 
thermodynamic information 
(compressibility, spin-susceptibility, phase diagrams 
and critical exponents) from density and density-fluctuation 
profiles of imbalanced mixtures of equal or unequal masses. 
We discussed continuum 
systems with and without trapping potentials. 
Using the local density approximation, 
we obtained expressions relating the local compressibility and local 
spin susceptibility to the local fluctuations in particle numbers.
Lastly, we applied our results to the case of 
population imbalanced Fermi systems of equal masses. 

\acknowledgments{We would like to thank the Army Research Office (Contract No. W911NF-09-1-0220) for support.}

\end{document}